\documentclass[apj,numberedappendix]{emulateapj}

\usepackage{natbib}
\usepackage{bm}		
\usepackage{bbm}	
\usepackage{mathptmx}	

\newcommand\arcsec{\mbox{$^{\prime\prime}$}}
\newcommand\arcmin{\mbox{$^{\prime}$}}

\newcommand{\Sigc}{\Sigma_{\rm crit}}
\newcommand{\zsrc}{z_{\rm src}}
\newcommand{\Ftot}{{\bf F}_{\rm tot}}

\begin{document}

\author{Neal Dalal\altaffilmark{1}} 
\affil{Institute for Advanced Study, Einstein Drive, Princeton NJ 08540}
\author{Joseph F.\ Hennawi and Paul Bode} 
\affil{Department of Astrophysical Sciences, Princeton University, 
Ivy Lane, Princeton NJ 08544}

\altaffiltext{1}{Hubble Fellow}

\title{Noise in strong lensing cosmography}

\begin{abstract}
Giant arcs in strong lensing galaxy clusters can provide a purely
geometric determination of cosmological parameters,
such as the dark energy density and equation of state. 
We investigate sources of noise in cosmography with giant arcs,
focusing in particular on errors induced by density fluctuations along
the line-of-sight, and errors caused by modeling uncertainties.  We
estimate parameter errors in two independent ways, first by developing
a Fisher matrix formalism for strong lensing parameters, and next by
directly ray-tracing through N-body simulations using a multi-plane
lensing code.
We show that for reasonable power spectra, density fluctuations from
large-scale structure produce $>100\%$ errors in cosmological
parameters derived from any single sightline, precluding the use of
individual clusters or ``golden lenses'' to derive accurate cosmological
constraints.  Modeling uncertainties similarly can lead to large
errors, and we show that the use of parametrized mass models in
fitting strong lensing clusters can significantly bias the inferred
cosmological parameters.  We lastly speculate on means by which these
errors may be corrected.
\end{abstract}


\maketitle

\section{Introduction}

There is strong observational evidence supporting the flat
$\Lambda$CDM cosmological model, in which the energy budget of the
universe is dominated by a poorly understood dark sector
\citep[e.g][]{bennett03,riess04}.  An important question is 
whether dark energy, which comprises at present
$\Omega_\Lambda\approx0.7$ of the critical density, is truly a
cosmological constant or instead a dynamical field with evolving
energy density.  Considerable effort has been directed towards
developing means of measuring the redshift evolution of dark energy,
and one of the most promising methods involves gravitational lensing
tomography.  The basic idea is to exploit the fact that the lensing
efficiency for a lens at distance $D_l$ and source at $D_s$ is
proportional to $D_l\,D_{ls}/D_s$.  By measuring the variation of
lensing strength with source redshift, one can measure distance ratios
as a function of redshift, and thereby constrain the background
cosmology.  Tomographic weak lensing is also sensitive to the growth
of structure, providing an additional handle on dark energy
\citep[e.g.][]{hu02}, although it may be possible to separate the
effects of growth and geometry in weak lensing tomography
\citep{jaintaylor03,zhang03} if a purely geometric determination of
the cosmology is desired.

This method may also be applied in the strong
lensing regime \citep[e.g.][]{bn92}.  \citet{golse02} discuss in
detail the cosmographic prospects for a sample of galaxy clusters,
each with multiple giant arcs at different redshifts, and
\citet{soucail04} apply this method to the galaxy cluster Abell 2218
to derive preliminary cosmological constraints.  Cluster surveys, 
such as the MACS survey \citep{ebeling01}, EDisCS survey \citep{ediscs},
or SDSS cluster lensing survey \citep{sdsslens},
and wide-area imaging surveys like the CFHT Legacy
Survey\footnote{http://www.cfht.hawaii.edu/Science/CFHLS}
or the RCS-2 survey are expected to find (and are finding!)
large numbers of arc-bearing galaxy clusters.  Frequently,
deep imaging reveals multiple arcs in the same cluster --
a spectacular example is provided by
Abell 1689 \citep{a1689}\footnote{see also
http://hubblesite.org/newscenter/newsdesk/archive/releases/2003/01/image/a}.
With this imminent explosion in strong lensing data, it is now
appropriate to consider realistic sources of noise in lensing
cosmography, with an aim towards determining optimal observational
strategies.  For example, is it better to devote large amounts of
telescope time to a few clusters, in the hopes of
finding dozens of arcs per cluster which completely constrain the mass
model, or is it better to survey a large number of clusters and study
only a few arcs per cluster?

In this paper, we study two principal sources of noise in strong
lensing cosmography.  First, we consider how uncertainties in the mass
modeling of the lensing cluster translate into errors in the derived
cosmological parameters.  Next, we investigate the effects of random
density fluctuations in the line of sight uncorrelated with the main
lensing cluster.  We find that density fluctuations can lead
to large errors in the inferred cosmology
along any given sightline.  This requires
the use of multiple sightlines; that is many lensing clusters must be
studied in order to suppress the noise associated with large-scale
structure.  With limited telescope time, this may also lead to
problems: the limited constraints provided by small numbers of arcs
per cluster require the use of parametric mass models in fitting the
lens data, and we show that such parametric models can produce biases
in the inferred cosmological parameters.

\section{Sources of noise}

\begin{figure}
\plotone{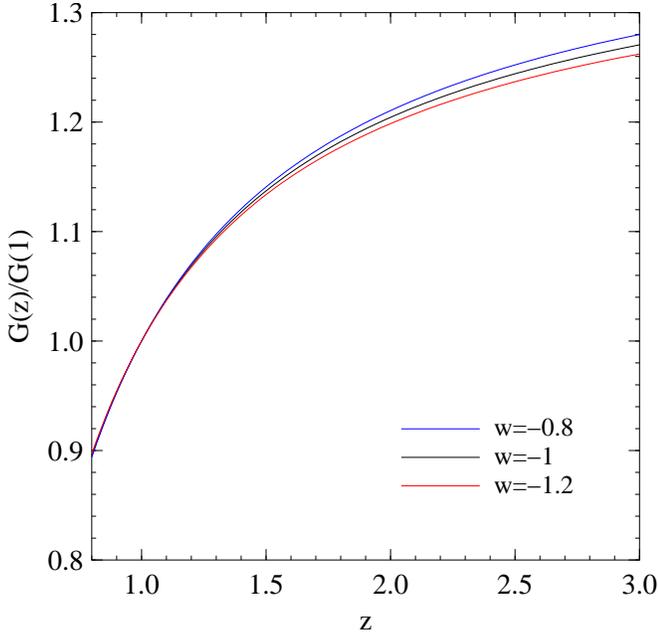}
\caption{Ratio of lensing efficiency, $G(z)/G(1)$, as a function
  of redshift.  The curves correspond to dark energy equation of state
  parameter $w=-0.8,\,-1,\,-1.2$ from top to bottom.
\label{ratio}}
\end{figure}

It is perhaps worth emphasizing the well-known point that
determining the properties of
dark energy using gravitational lensing tomography is quite difficult,
because variation of dark energy parameters produces small effects on the
lensing observables.  For tomographic lensing, the basic observable is
the variation of lensing efficiency $G\propto D_l\,D_{ls}/D_s$ with
source redshift.  We do not measure the absolute lensing efficiency,
only the ratio of efficiencies at different redshifts.  In
figure~\ref{ratio} we plot the ratio $G(z)/G(1)$ for lens
redshift $z_l=0.3$, as a function of source redshift $z$ for several
values of $w$.  As can be seen, in order to produce interesting
constraints on $w$ we must be able to measure the lensing efficiency
at the percent level, which is quite challenging.
Weak lensing surveys aim to overcome this challenge using brute force,
by surveying large fractions of the sky and measuring ellipticities of
huge numbers of galaxies.  Because strong lensing events are rare, it
would be desirable to focus instead on a small number of objects and
study them well.  The fact that the signal is so small, however, means
that strong lensing cosmography will be subject to a wide variety of
error sources.

\subsection{Uncertainties in mass modeling}
\label{obserr}

Uncertainties in mass modeling have long impeded the cosmographic
applications of strong lenses.  For example, the main systematic
uncertainty in determination of the Hubble constant $H_0$ using lens
time delays continues to be the radial density profile of the lens
\citep[e.g.][]{kochanek02}. Similarly, uncertainties in
the mass modeling of lensing clusters also induce errors in the
inferred cosmology.  The reason is easy to understand -- variations in
the dark energy evolution, which change $\Sigc(z)$,  
can be compensated by adjusting the model
density profile $\Sigma(R)$ to hold fixed the observables, which
depend only upon the ratio $\Sigma/\Sigc$.    

In the appendix, we calculate the parameter errors expected given a
lens model and a specified level of observational errors.  As shown in
Appendix~\ref{app:cov}, we can write the Fisher matrix as a sum over
Fisher matrices for individual sources,
\begin{equation}
\Ftot = \int d\zsrc \, n(\zsrc) \int d^2\!u\, {\bf F}(\bm u, \zsrc).
\end{equation}
Here, $n(\zsrc)$ describes the number density of sources as a function
of redshift $\zsrc$, and the integral over source position $\bm u$ is
restricted to the strong lensing region at each redshift.
To illustrate the level of errors expected, let us consider strong
lensing by two classes of models: an ellipsoidal NFW \citep{nfw}
model, and the power-law ellipsoid model of \citet{barkana98}.
We additionally allow parameters for external shear, along with the
cosmological parameters (here taken to be only $\Omega_M$ and $w$),
all of which must be simultaneously determined from the image data.  
Assume that there are $N$ multiply imaged sources used to constrain
the model, drawn from a population following a redshift distribution 
$n(z)\propto z^2\exp(-(z/z_w)^2)$ \citep{hujain04} over a range
$0.8<\zsrc<4$, with median redshift $z_{\rm med}=1.5$.  
Unless stated otherwise, we will assume
for the underlying cosmology a flat
$\Lambda$CDM model with WMAP parameters \citep{bennett03}: 
$\Omega_M=0.27$, $\Omega_\Lambda=0.73$, $\Omega_Bh^2=0.024$, 
$h=0.72$, $n=1$, $\sigma_8=0.9$.

We first consider cosmological constraints from power-law ellipsoids.  
The fiducial parameter values we adopt for this lens model are:
Einstein radius $b=10\arcsec$ (for $\zsrc=4$), axis ratio $q=0.7$, 
isothermal profile, 
vanishing core radius, no external shear, and $\Omega_M=0.27$, $w=-1$.  
For concreteness, we take observational errors of $\sigma=0.1\arcsec$,
appropriate for {\em Hubble Space Telescope} imaging.  
Ignoring the degeneracies between the cosmological parameters
$(\Omega_M,w)$ and the lens parameters we would
obtain errors $\sigma_\Omega=0.07N^{-1/2}$ and $\sigma_w=0.1N^{-1/2}$
with $N$ arcs.  However, marginalizing over the lens parameters, these
errors explode up to $\sigma_\Omega=3.6N^{-1/2}$ and
$\sigma_w=5N^{-1/2}$.  

Next we consider the NFW model.  One qualitative difference between
this model and the singular isothermal ellipsoid is that the NFW
profile produces a central image at small radii.  This increases the
dynamic range of the lensing constraints, and helps break degeneracies
between the cosmology and the mass profile.  Using fiducial NFW
parameters of $M_{\rm vir}=5\times10^{14}h^{-1}M_\odot$, $c_{\rm vir}=5$,
axis ratio $q=0.7$ and no external shear, we find improved
constraints: $\sigma_\Omega=0.5N^{-1/2}$, $\sigma_w=0.7N^{-1/2}$.  The
improvement can be explained in part by the extra constraints from the
central image, and in part by the smaller number of parameters in the
NFW model compared to the power-law ellipsoid.  

In reality, central images are rarely detected in strong lenses, and
are strongly affected by the properties of the central cD galaxy in
the cluster.  Additionally, 
we expect real clusters to exhibit a broader range of density profiles
than can be accommodated by the NFW model, due to, for
example, the
effect of baryons.  We therefore expect the errors in real clusters to 
be significantly larger than the errors we have computed for NFW
ellipsoids.   Note that these errors are only those caused
by propagating the observational uncertainties; below we discuss
additional sources of error.  Already, though, these error levels 
would require large numbers of arcs in a given cluster
for useful constraints to be placed on dark energy.

\subsection{Systematic errors in the mass model}

One reason for the large errors found above is that we employed models with 
many parameters, producing degeneracies with the cosmological
parameters.  This requires large numbers of arcs in order to constrain
the model parameters sufficiently.  Had we taken a less flexible
model, we would not have found as large errors on the cosmological
parameters.  Unfortunately, real clusters are expected to exhibit a
broad range of behavior, as found in N-body simulations.  If the
assumed mass 
model does not fully capture the range of behavior of the real density
profile, then the derived cosmological parameters will be biased;
that is,
the fit can be improved by adjusting the cosmology to make the convergence
$\kappa(R,z_s)$ a better fit to the density profile than variations of
the mass model $\Sigma(R)$ can accomplish alone.

\begin{figure}
\plotone{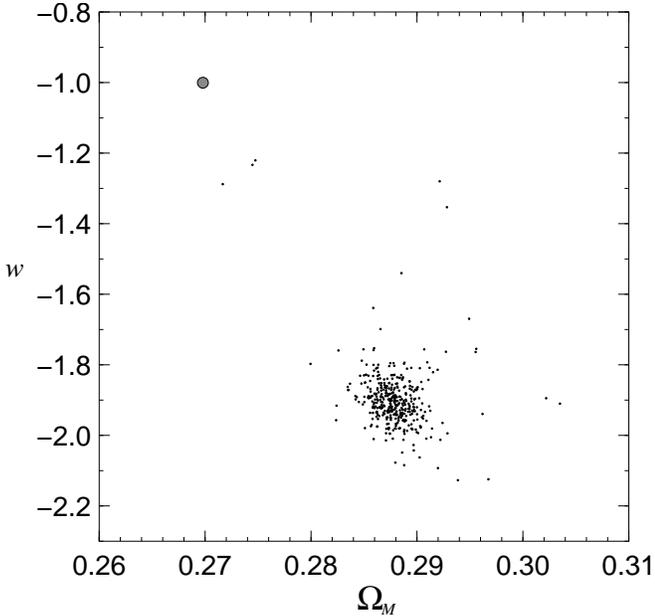}
\caption{Fits of NFW lenses to softened power-law
  ellipsoidal models.  Each point corresponds to the best-fit values of
  $\Omega_M$ and $w$ derived by modeling sets of 50 randomly generated
  arcs at 5 different redshifts.  The input cosmology had
  $\Omega_M=0.27$ and $w=-1$, as depicted by the filled circle.
  \label{modbias}
}
\end{figure}

This bias is illustrated in figure \ref{modbias}.  The figure shows
results of a Monte Carlo calculation, in which artificial arcs are
generated using ellipsoidal NFW profiles, and fitted using softened
power-law ellipsoidal models \citep{barkana98}, which can mimic a wide
range of radial profiles over the relevant radii of interest for
strong lensing.  A CMB prior on the $\Omega_M-w$ plane has also been
imposed: the distance to the last scattering surface at $z=1089$ is
required to be $d_A = 14\pm0.2$ Gpc for $h=0.71$ \citep{bennett03}; 
the degenerate direction is along decreasing $w$ (i.e.\ more negative)
as $\Omega_M$ is increased.  
For each realization, we generate 10 arcs each at redshifts of 
$\zsrc=0.8$, 1.6, 2.4, 3.2 and 4 using the same NFW model, and
fit all 50 arcs using the power-law ellipsoid model, allowing
$\Omega_M$ and $w$ to vary in the fit.  We plot only the best-fitting
$\Omega_M$ and $w$ for each realization; although the confidence
region for each realization is large, we can detect parameter biases
by examining the best-fitting models.  
Clearly, the cosmological parameters derived from multiple arcs can be
biased at significant levels (in this example, $\Delta w\sim 1$).  The
sign and direction of the bias depend upon the profile shapes for the
cluster and the model, and could be different from cluster to cluster
if there is significant variation among cluster profiles.  The bias
can be suppressed by using more flexible models in the fits.
Unfortunately, as discussed above, extra degrees of freedom in the
mass model come at the expense of weaker constraints on cosmology.

\subsection{Projection noise}

In addition to the modeling uncertainties discussed above, another
source of noise in strong lensing cosmography is large-scale
structure.  Fluctuations in the density along the line of sight, if
uncorrected, produce errors in the derived cosmological parameters.  
Again, this effect is easy to understand -- density fluctuations at
redshifts different than the primary lens redshift have different
lensing efficiencies than the primary lens, producing effects varying
with source redshift which appear similar to variations in the
background cosmology.
Naively, one might expect the parameter errors to be small -- the
density fluctuations are at the $\lesssim 1\%$ level, and geometric
factors suppress the effects of fluctuations at redshifts very
different from the cluster redshift.  

In the appendix, we compute the errors expected in the cosmological
parameters caused by density fluctuations along the line of sight.  We
use the same fiducial lens models as in section \S\ref{obserr}, and
estimate the line-of-sight density fluctuations by projecting the
matter power spectrum using the \citet{limber} approximation,
\begin{equation}
\delta\Sigma^2=\frac{9\pi}{4}\Omega_M^2 \Sigma_0^2 
\int_{z_1}^{z_2} \! dz \frac{H_0}{H} 
\int \frac{dk}{k} \frac{\Delta^2(k,z)}{k\,r_H} W_2^2(k r \theta),
\end{equation}
with $\Sigma_0=c H_0/4\pi G$, $r_H = c/H_0$, 
$\Delta^2=k^3 P(k)/2\pi^2$, and $W_2(x) = 2 J_1(x)/x$, with a
smoothing scale $\theta=30\arcsec$.  We smooth the power spectrum in
order to select only those fluctuations that are coherent over the
strong lensing region; smaller scale fluctuations could be suppressed
by observing large numbers of arcs in the cluster, and by modeling of
individual galaxies as substructure in the mass map.  
We use the nonlinear power spectrum of \citet{pd96} and the
approximate transfer functions of \citet{eisensteinhu99}, for a flat
$\Lambda$CDM cosmology with WMAP parameters.
We used 10 lens planes to describe the
line-of-sight density fluctuations, assuming that the shear variance
equals the convergence variance (and assuming that the shear is
randomly oriented in each plane).  When we include these line-of-sight
fluctuations and single-plane lens modeling, as described in the
appendix, we obtain errors of $\sigma_\Omega\simeq 0.34$ and 
$\sigma_w\simeq 0.32$ for the NFW model, and $\sigma_\Omega\simeq1$
and $\sigma_w\simeq 2$ for the singular power-law ellipsoid model. 
Note that these errors do not diminish when large numbers of arcs are
observed per cluster; the projection errors are suppressed only by
viewing multiple independent sightlines.  

\section{Combined error forecast}

In this section, we combine all of the above sources of noise to
compute the expected error budget for strong lensing cosmography with
giant arcs.  We ray-trace through N-body simulations using code
described in \citet{dalal04}, modified slightly to perform multi-plane
lensing, and construct mock catalogues of giant
arcs.  Then we fit the resulting arc data using ellipsoidal NFW mass
models, optimizing both the lens parameters and the cosmological
parameters $\Omega_M$ and $w$.  One advantage of using N-body
simulations is that we can account for the effects of the highly skewed
distribution of density fluctuations in the nonlinear
regime; in principle this non-Gaussianity could lead to additional 
biases in the cosmological parameters which would not be revealed by 
the Limber approximation used above.

The N-body simulation we use is described in \citet{wambsganss03}.
Briefly, the TPM code \citep{tpm} was used to simulate a box with
side length $L=320 \,h^{-1}$Mpc in a flat $\Lambda$CDM cosmology with
$\Omega_M=0.3$, $\Omega_\Lambda=0.7$, $h=0.7$, $n_s=1$ and
$\sigma_8=0.95$; this is broadly consistent with WMAP parameters
\citep{bennett03}.  A total of $N=1024^3$ particles were used in the
simulation.  We then sliced through the simulation volume to produce
lens planes spaced every $160\,h^{-1}$Mpc with angular size
roughly 20\arcmin, so that the physical side length of the projected
planes telescopes with increasing distance from the observer at
$z=0$.  A total of 243 pairs of planes were produced for each of 19 redshifts
between $z=0$ and 6.37.  We subtracted the cosmic mean density from
each plane and then tiled the lightcone by randomly selecting a
pair of
planes for each redshift, for a total of 38 lens planes per realization.

\begin{figure}
\plottwo{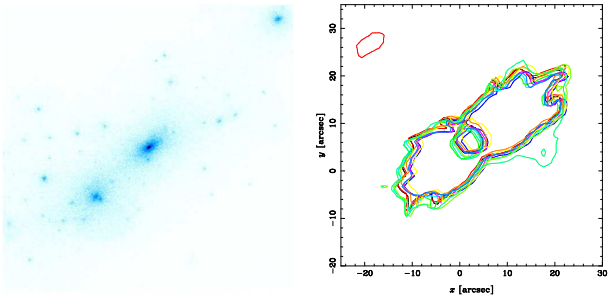}{f3b.eps}
\caption{(Left) A simulated cluster at $z=0.29$ used to generate arcs.
  The cluster has a virial mass of $M=10^{14.5}\,h^{-1} M_\odot$ and
  $c_{\rm vir}=7$.  The side length is $2\,h^{-1}$ Mpc comoving.
  (Right) The $\zsrc=5$ critical lines for this cluster.  The black
  curve corresponds to the critical lines for no intervening lens
  planes, and the colored curves are the critical lines for 10 different
  realizations of the intervening matter.
\label{clus}}
\end{figure}

In order to ensure that strong lensing occurs, we insert an extra lens
plane containing a massive cluster, along with all the surrounding
matter within $5 h^{-1}$Mpc.  An example of one of the clusters used
is shown in fig.~\ref{clus}.  We have used only clusters with regular,
relaxed cores, since clusters with multiple massive components require
more sophisticated modeling than is feasible using automated routines.
We first generate arcs at multiple
source planes using only the cluster lens, neglecting the other lens
planes tiling the light cone.  This amounts to assuming that matter is
unclustered at redshifts different than the cluster redshift.  
We fit these arcs using the projected NFW
model, holding fixed the cosmological parameters, to create a
starting model from which subsequent fits are started.  Next, we tile
the lightcone by randomly selecting slices as described above, and
recompute the ray-trace.  We again generate sets of strongly lensed
sources, and model the mock data sets using single-plane mass models,
starting from the models which best fit the single-plane lens data, 
optimizing over both lens parameters and the cosmological parameters
$\Omega_M$ and $w$.  In general, the lensing properties of the
clusters remain quite similar with and without the extra lens planes.
For example, the right panel of fig.~\ref{clus} shows the critical
curves for the cluster shown in the left panel, for sources at $\zsrc=5$,
and it is apparent that the critical curves are only slightly affected
by the LOS density fluctuations.  Similarly, the total lensing cross
sections and optical depth are negligibly affected by inclusion of the
LOS material.

\begin{figure}
\plotone{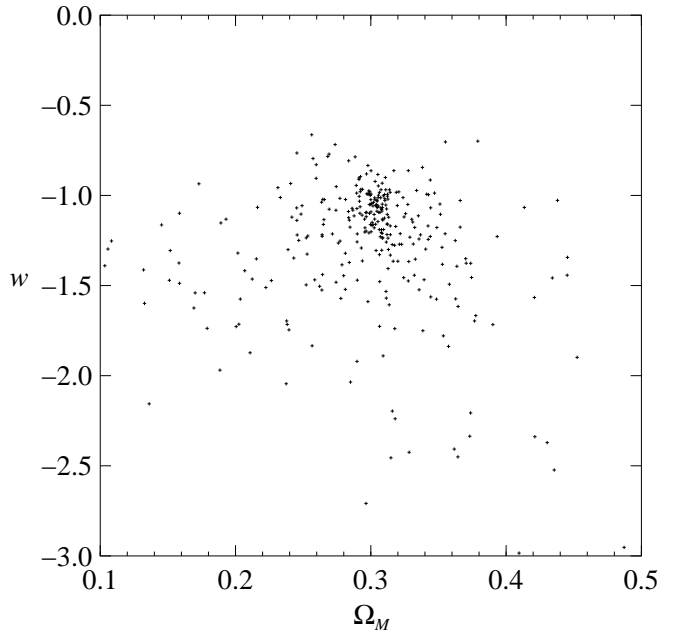}
\caption{Cosmological parameters derived from fitting sets of 50
  strongly lensed sources between redshifts $\zsrc=0.8$ to 5,
  generated by ray-tracing through N-body simulations with full
  lightcone tiling.  Each point corresponds to a realization of the
  intervening planes and lensed images.  An input cosmology of
  $\Omega_M=0.3$, $w=-1$ was used to generate the lensed images.
\label{simfig}}
\end{figure}

Nevertheless, considerable errors in the derived cosmological parameters
can arise from projection noise as
well as deviations of the cluster profiles from the simple NFW form.  
Figure~\ref{simfig} shows results from our ray-tracing simulations.
As is apparent from the figure, quite large, biased errors can arise
from the combination of effects discussed above.  The 95\% confidence
regions for the two fitted parameters are $0.13<\Omega_M<0.45$ and
$-3.3<w<-0.7$ with mean values of $\langle\Omega_M\rangle=0.3$ and
$\langle w\rangle=-1.3$, for input values of $\Omega_M=0.3$ and
$w=-1$.  

\section{Cleaning out the noise}

The sources of noise discussed above pose serious challenges to
the use of giant arcs for cosmography, however they are not
necessarily insurmountable.  Modeling uncertainties can be removed by
observing large numbers of arcs in each cluster, so that a
nonparametric, unbiased reconstruction of the mass profile and
cosmology may be performed.  This requires high resolution and deep
imaging, along with redshifts for the arcs.

Projection noise appears more recalcitrant.  Directly measuring the
intervening mass fluctuations with weak lensing
\citep[e.g.][]{taylor01} is difficult, in
part because the main lens cluster must be carefully subtracted, but
also because weak lensing can only measure the fluctuations on scales
larger than a few arcminutes, while the strong lensing region is
sensitive to power down to sub-arcminute scales \citep{dalal03}.
Another possible means of cleaning out the projection noise would be to
reconstruct the density and shear fluctuations along the
line-of-sight using the observed galaxies in the field.  This is not
unreasonable, since we know that most of the power on the relevant
$\sim$arcminute scales is produced by halos \citep{seljak00}.
However, previous attempts at such reconstructions have not been entirely
successful, with different groups finding quite different shear and
convergence over the same field 
\citep[e.g.][]{lewisibata01,riess01,goobar01,benitez02}.  

The main difficulty with predicting the lensing convergence and shear
fields from the field galaxies is that galaxies and mass are not
perfectly correlated.  Recall that we must be able to suppress the
projection noise by at least 1-2 orders of magnitude in order to
derive useful cosmological constraints from multiple arcs in a single
cluster.  Unfortunately, the cross-correlation between galaxies and
mass, $r=P_{gm}/(P_{gg} P_{mm})^{1/2}$, is expected to be less than
unity, typically of order $r\sim 0.5$ on the relevant 
scales.\footnote{Note that observational estimates
\citep[e.g.][]{hoekstra01,sheldon04,sdssbias} typically quote $r>1$ on
small scales because the shot noise is subtracted from the galaxy power
spectrum $P_{gg}.$}  This would 
appear to preclude a sufficiently accurate reconstruction.  Another
way to understand this is as follows.  Reconstructing the convergence
and shear arising from individual galaxies in the field involves
associating halos of mass $M$ with galaxies of luminosity $L$.
Unfortunately, the relation between $M$ and $L$ is not one-to-one.
For example, a galaxy like the Milky Way may be the central galaxy in
halo of mass $10^{12}M_\odot$, or it may be a member of a group an
order of magnitude larger in mass.  In the halo occupation
distribution model of \citet{kravtsov03}, there is roughly a factor of
$3-5$ uncertainty in the mass of a halo hosting a given galaxy, which
directly translates into the uncertainty in the convergence and shear
associated with the galaxy.  Even if central galaxies may be
distinguished from satellite galaxies, there is still an unknown
scatter in the relation between halo mass and central galaxy
luminosity \citep{iro04,kev04}.  In addition, halo triaxiality and
substructure will affect the convergence and shear at the level of
interest.

One crude test of the reconstruction method may be possible using
extant cosmic shear data.  Using the observed foreground galaxies in
the field, one can make a prediction for the cosmic shear as a
function of position, which can be checked against the values actually
observed.  This unfortunately cannot test the fidelity of the
reconstruction on the required $\sim 10\arcsec$ scales, since shot
noise in the source galaxies typically limits the resolution of shear
maps to arcminute scales.  While this cannot fully validate the
method, a test along these lines would be a worthwhile indicator of
whether significant observational efforts should be devoted towards
the use of giant arcs to constrain geometry.

\section{Summary}

We have quantified various sources of noise expected in strong lensing
cosmography.  The use of multiple arcs in galaxy clusters holds great
promise towards constraining the properties of dark energy,
however significant errors can arise
from line-of-sight projections and modeling uncertainties.  Removing
the projection noise to the necessary level appears difficult, since
reconstruction of the mass density along the line of sight is hampered
by the fact that mass and light are not perfectly correlated.  If
projection noise cannot be adequately cleaned out, then we cannot use
``golden lenses'', but must instead average over many strong lensing
clusters in order to derive useful cosmological constraints.  To avoid
biases in the inferred cosmological parameters, highly flexible or
nonparametric mass models must be used to describe the cluster.  In
the absence of any knowledge of cluster density profiles, 
non-parametric models require large numbers of arcs to be observed in
each cluster to sufficiently constrain the mass model.  
However, in principle one could use N-body simulations to impose prior
constraints on cluster profiles.  Using priors from N-body simulations
would not be useful for individual systems, because of the large scatter
in profiles among individual clusters.
However, if simulated clusters do provide a reasonable description of
the population of real galaxy clusters, then priors derived from
simulations may allow useful cosmological constraints to be derived
from large ensembles of lensing clusters.  \citet{meneghetti04} make
preliminary investigations into this approach.  
It may also be worthwhile to investigate whether
additional information, such as X-ray temperature profiles of lensing
clusters, can be used to improve the cosmological constraints derived
from giant arcs.

\acknowledgments{
We thank B.\ Frye for helpful discussions, and thank B.\ Jain
and J.\ Ostriker for comments on an earlier version of this paper.
N.\ D.\ acknowledges the support of NASA through Hubble Fellowship
grant \#HST-HF-01148.01-A awarded by the Space Telescope Science
Institute, which is operated by the Association of Universities for
Research in Astronomy, Inc., for NASA, under contract NAS 5-26555.
J.\ F.\ H.\ is supported by a Proctor Fellowship granted by Princeton
University. Computations were performed on NSF supported systems
at the Pittsburgh Supercomputing Center and the National Computational 
Science Alliance; also at Princeton University on 
facilities provided by NSF grant AST-0216105.
}

\appendix

\section{Covariance matrix} \label{app:cov}

Consider a model for a lens system with $n_{\rm obs}$ observables
${\bm O}$ and $n_p$ parameters ${\bm p}$ describing
the lens (and possibly cosmology as well).  Assume that the penalty
function may be written 
$\chi^2=\bm{\delta O}\cdot{\bf V}\cdot\bm{\delta O}$,
where ${\bf V}^{-1} = \langle{\bm{\delta O\,\delta O}}\rangle$ describes
the covariance of the observational errors.  For each lensed source,
there are additionally $m$ parameters, with $m=2$ if only the image
positions are being modeled, or $m=3$ if the fluxes are modeled as
well.  The $m\,n_{\rm src}$ source parameters are nuisance parameters,
over which we marginalize to obtain the $n_p\times n_p$ covariance
matrix and its inverse, the Fisher matrix ${\bf F}$.  If 
${\bf D}=\partial{\bm O}/\partial {\bm p}$, then propagating errors we
have ${\bf F}={\bf D}^{\sf T}\cdot{\bf V}\cdot{\bf D}$.  For a single
source, the (unmarginalized) Fisher matrix takes the form
\begin{equation}
\bordermatrix{n_p\mbox{\large\{} &
\overbrace{\bf A}^{\displaystyle n_p} & 
\overbrace{\bf B}^{\displaystyle m} \cr
m\,\mbox{\large\{} & {\bf B}^{\sf T} & {\bf C}\cr
},
\label{onesrc}
\end{equation}
where the $n_p\times n_p$ block ${\bf A}$ corresponds to the lens
parameters, the $m\times m$ block ${\bf C}$ corresponds to the
source parameters, and the $n_p\times m$ block ${\bf B}$ is the cross
term. Marginalizing over the source parameters, the resulting Fisher
matrix becomes 
${\bf F}={\bf A}-{\bf B}\cdot{\bf C}^{-1}\cdot{\bf B}^{\sf T}$.

For multiple sources, the full (unmarginalized) Fisher matrix has the
form 
\begin{equation}
{\bf F}_u = 
\left(
\begin{array}{ccccc}
\sum_i{\bf A}_i & {\bf B}_1 & {\bf B}_2 & {\bf B}_3 & \ldots \\
{\bf B}_1^{\sf T} & {\bf C}_1 & 0 & 0 & \ldots\\
{\bf B}_2^{\sf T} & 0 & {\bf C}_2 & 0 & \\
{\bf B}_3^{\sf T} & 0 & 0 & {\bf C}_3 & \\
\vdots & \vdots & & & \ddots 
\end{array}
\right).
\end{equation}
Here, ${\bf A}_i$, ${\bf B}_i$, and ${\bf C}_i$ are the components of
the single-source Fisher matrices (eqn.~\ref{onesrc}) for source $i$. 
We wish to invert ${\bf F}_u$ to obtain the parameter covariance
matrix.  We can do so by noting that ${\bf F}_u$ is diagonalized by
the transformation
\begin{equation}
{\bf U}\cdot{\bf F}_u\cdot{\bf U}^{\sf T}=
\left(\begin{array}{cccc}
\Ftot & 0 & 0 & \ldots \\
0 & {\bf C}_1 & 0 & \ldots\\
0 & 0 & {\bf C}_2 & \\
\vdots & \vdots & & \ddots 
\end{array}
\right),
\end{equation}
where 
\begin{equation}
{\bf U}=
\left(\begin{array}{cc}
\mathbbm{1} & 
\begin{array}{ccc}
-{\bf B_1}{\bf C}_1^{-1} & -{\bf B_2}{\bf C}_2^{-1} & \ldots \end{array} \\
\begin{array}{c}
0 \\ 0 \\ \vdots \end{array}
& {\mbox{\Huge $\mathbbm{1}$}}
\end{array}
\right),
\end{equation}
and $\Ftot=\sum_i{\bf F}_i
=\sum_i {\bf A}_i - {\bf B}_i\cdot{\bf C}_i^{-1}\cdot{\bf B}_i^{\sf T}$.
From this, it is easy to see that 
\begin{eqnarray}\label{covar}
{\bf F}_u^{\,-1} &=& {\bf U}^{\sf T}\cdot
\left(\begin{array}{cccc}
\Ftot^{\,-1} & 0 & 0 & \ldots \\
0 & {\bf C}_1^{-1} & 0 & \ldots\\
0 & 0 & {\bf C}_2^{-1} & \\
\vdots & \vdots & & \ddots 
\end{array}
\right)\cdot {\bf U} \\\nonumber
&=&\left(\begin{array}{cccc}
\Ftot^{\,-1}\qquad & -\Ftot^{-1}{\bf B}_1{\bf C}_1^{-1} &  
-\Ftot^{-1}{\bf B}_2{\bf C}_2^{-1} & \ldots \\
-{\bf C}_1^{-1}{\bf B}_1^{\sf T}\Ftot^{-1}\qquad & {\bf C}_1^{-1} + 
{\bf C}_1^{-1}{\bf B}_1^{\sf T}\Ftot^{-1}{\bf B}_1{\bf C}_1^{-1} & 
{\bf C}_1^{-1}{\bf B}_1^{\sf T}\Ftot^{-1}{\bf B}_2{\bf C}_2^{-1} & \ldots\\
-{\bf C}_2^{-1}{\bf B}_2^{\sf T}\Ftot^{-1}\qquad & 
{\bf C}_2^{-1}{\bf B}_2^{\sf T}\Ftot^{-1}{\bf B}_1{\bf C}_1^{-1} & 
{\bf C}_2^{-1} + {\bf C}_2^{-1}{\bf B}_2^{\sf T}\Ftot^{-1}{\bf B}_2{\bf C}_2^{-1} & \\
\vdots & \vdots & & \ddots 
\end{array}
\right)
\end{eqnarray}
The covariance of the $n_p$ lens parameters is given by the upper left
$n_p\times n_p$ block $\Ftot^{\,-1}$.
Therefore, the full (many-source) marginalized Fisher matrix is simply
the sum of the marginalized single-source Fisher matrices, as we might
have expected.

\section{Large-scale structure}

Density fluctuations along the line of sight will perturb the lensed
images, inducing changes in the derived best-fitting lens parameters.
\citet{dk02} compute the perturbations to the parameters to first
order in the density and shear perturbations, which is sufficient for
our purposes (recall that the rms convergence fluctuations are
$\langle\kappa^2\rangle^{1/2}\simeq 1\%$).  The only minor
modification here is that multi-plane lens factors must be included
\citep{sef,keeton03}.  

Let us parametrize the line-of-sight fluctuations by projecting the
density and shear onto additional lens planes.  Each plane has its own
surface density perturbation $\delta\Sigma$ (measured relative to mean
density) and 
(2-component) tidal shear $\delta\Gamma$, so that for $N$
additional planes we have $n_q=3N$ variables, which we denote by
$n_q$ element $\bm q$.  Just as shifts in the lens parameters
$\Delta\bm p$ produce shifts in the observables 
$(\partial\bm O/\partial\bm p)\Delta\bm p$, the line of sight
fluctuations produce shifts $(\partial\bm O/\partial\bm q)\Delta\bm q$. 
Minimizing $\chi^2$, this implies that the LOS fluctuations induce
parameter shifts 
\begin{eqnarray}
\Delta\bm p &=& {\bf F}_u^{\,-1}\cdot
\left(\frac{\partial\bm O}{\partial\bm p}\right)^{\sf T}\cdot{\bf V}
\cdot\frac{\partial\bm O}{\partial\bm q}\cdot \Delta\bm q \nonumber\\
&\equiv& {\bf M}\cdot \Delta\bm q. \label{M}
\end{eqnarray}
We are mainly interested in the induced shifts in the lens parameters
and cosmological parameters, not the nuisance parameters describing
the sources.  Let us write the first $n_p$ elements of the full
parameter array $\bm p$ as ${\bm p}_{\rm lens}$, and the $m$ nuisance
parameters for each of the sources as $\bm u$.  
Using the above expression for ${\bf F}_u$ (eqn.~\ref{covar}), we can
write the first $n_p$ rows of eqn.~\ref{M} as a sum over independent
terms for each source,
\begin{equation}
{\bf M}_{1\ldots n_p} = \Ftot^{\,-1} \sum_i^{n_{\rm src}} \left[
\left(\frac{\partial\bm O_i}{\partial\bm p_{\rm lens}}\right)^{\sf T}
-{\bf B}_i {\bf C}_i^{-1}
\left(\frac{\partial\bm O}{\partial\bm u}\right)_i^{\sf T}\right]
\cdot{\bf V}\cdot\frac{\partial\bm O_i}{\partial\bm q}.
\end{equation}
Given our expression for ${\bf M}$, 
the parameter covariance caused by LOS fluctuations is
$\langle\Delta p_i\Delta p_j\rangle = M_{i\alpha}
\langle\Delta q_\alpha\Delta q_\beta\rangle M_{j\beta}$.


\end{document}